\documentclass{aa}

\usepackage{graphicx}
\usepackage{txfonts}
\usepackage{color}

\begin{document}

\title{Preprocessing in small groups:\\ Three simulated galaxies interacting prior to cluster infall
}

\author{Ewa L. {\L}okas
}

\institute{Nicolaus Copernicus Astronomical Center, Polish Academy of Sciences,
Bartycka 18, 00-716 Warsaw, Poland\\
\email{lokas@camk.edu.pl}}

\titlerunning{Preprocessing in small groups}
\authorrunning{E. L. {\L}okas}


\abstract{
The formation of galaxy clusters is a complicated process that probably involves the accretion of galaxies in groups,
as observed in nearby clusters, such as Virgo and Fornax. The members of the groups undergo "preprocessing" prior to
cluster infall, which affects their stellar populations and morphology. In this paper I present an
extreme example of such an accretion event selected from the IllustrisTNG100 simulation. The group, composed of three
full-sized disky galaxies and a number of smaller satellites, is accreted early, with the first pericenter around the
cluster at redshift $z=1.3$. Before the infall, the three galaxies interact strongly in pairs within the group, which
produces tidally induced bars in the two more massive ones. The interactions also lead to mass exchange and
trigger some star formation activity resulting in temporary rejuvenation of their stellar populations. After infall,
they all undergo seven pericenter passages around the cluster, experiencing strong mass loss in the dark matter and gas
components, as well as reddening of the stellar populations. Their tidally induced bars are, however, preserved and
even enhanced probably due to the loss of gas via ram-pressure stripping in the intracluster medium. The study
demonstrates that group accretion can happen very early in cluster formation and proposes another scenario for the
formation of tidally induced bars.}


\maketitle

\section{Introduction}

According to the theory of hierarchical structure formation in the Universe dominated by cold dark matter, galaxy
clusters are the biggest gravitationally bound and virialized objects that have formed as of the present time. The
formation of galaxy clusters is a complicated process that is not yet fully understood. They do not form by accreting
galaxies one by one, nor do they collapse and virialize as a large region containing many galaxies over a short
timescale. Instead, many galaxies are accreted in groups, and while belonging to the groups they undergo
"preprocessing," which can include morphological, dynamical, and stellar population changes taking place before
the actual infall into the cluster and strong evolution inside it \citep{Fujita2004, Cortese2006, Haines2015}.

Cases of groups being accreted onto clusters have been studied observationally mostly in nearby clusters. For example,
the nearby Fornax cluster is known to possess a neighboring Fornax A group and \citet{Su2021} have recently studied the
signs of preprocessing between the Fornax main cluster and the group using the Fornax Deep Survey. \citet{Chung2021}
probed chemical preprocessing in star-forming dwarf galaxies in filamentary structures that are physically connected
to the Virgo cluster, while \citet{Bidaran2022} studied the effect of preprocessing on the stellar populations of
dwarf elliptical galaxies accreted onto the Virgo cluster $2-3$ Gyr ago as members of a massive galaxy group.
\citet{Sengupta2022} studied the more distant Abell 1882 at redshift $z=0.14$, a massive system including
several comparably rich substructures surrounded by filaments, and they found that quenching of star formation occurs in the
infall regions of these structures even before the galaxies enter the denser group environment. \citet{Sarron2019}
identified cosmic filaments around even more distant clusters at redshifts $0.15 < z < 0.7$ and found galaxy-type
gradients and other clues for preprocessing in such filaments that could be due to small galaxy groups.

The possibility of detection of subtle features that may result from preprocessing in groups is naturally limited in
higher redshift objects, so the early stages of cluster formation are presently beyond the reach of observations.
However, they can be studied in cosmological simulations that allow one to follow the evolution of galaxies from the early
stages with sufficient resolution to identify features that could be the imprint from preprocessing in groups. After
the first conflicting results from simulations concerning the role of groups in cluster formation \citep{Berrier2009,
McGee2009}, the consensus has been that the accretion in groups plays an important role in this process.

For example, \citet{DeLucia2012} used the results of the Millennium Simulation \citep{Springel2005} aided by
semi-analytic models and found that large fractions of group and cluster galaxies have been preprocessed as satellites
of groups with mass $\sim 10^{13}$ M$_\odot$. \citet{Benavides2020} studied ten massive clusters from the cosmological
simulation Illustris and determined that on average about 40\% of galaxies surviving at $z = 0$ were
accreted as part of groups and did not infall directly from the field. Most recently, \citet{Haggar2023} studied a
large sample of galaxy clusters from The Three Hundred project and more than a thousand groups passing through a
cluster to find that half of group galaxies become gravitationally unbound from the group by the first pericenter and
most groups quickly mix with the cluster satellite population.

Very promising tools to address these issues were provided by the simulations from the IllustrisTNG project
\citep{Springel2018, Marinacci2018, Naiman2018, Nelson2018, Pillepich2018}, where the evolution of galaxies was
followed by solving for the gravity and magnetohydrodynamics, and applying subgrid physics in the form of prescriptions
for star formation, galactic winds, magnetic fields, and the feedback from the active galactic nuclei (AGN). Numerous
analyses of the properties of galaxies that formed in these simulations have demonstrated that they are able to
reproduce the main features of galaxy populations found in observations. In particular, they have been successful in
reproducing many properties of galaxy clusters and the processes involved in galaxy evolution inside them
\citep{Pillepich2018, Gupta2018, Barnes2018, Yun2019, Sales2020, Joshi2020, Lokas2020b, Lokas2023, Dacunha2022}.

In this paper I present an extreme case of a small group of galaxies infalling into a cluster very early in
terms of the cosmological timescale, selected from IllustrisTNG simulations. The galaxies were found to interact
strongly during the infall, which led to the formation of tidally induced bars, a phenomenon well described in the
literature \citep{Noguchi1987, Gerin1990, Lokas2018, Peschken2019}. The bars were then enhanced and preserved during
the evolution in the cluster even though the galaxies followed quite tight orbits with many pericenter passages.
Although the emphasis here is on the morphological evolution of the galaxies, I also discuss some properties of their
stellar populations, such as the color and star formation rate (SFR).

\section{Results}

\subsection{Identification of the group}

The group was identified in the IllustrisTNG100-1 run following an earlier study of bar-like galaxies in
IllustrisTNG \citep{Lokas2021}. In that study I described a sample of galaxies with total
stellar masses greater than $10^{10}$ M$_\odot$ selected based on the single condition of a low enough ratio of the
intermediate to long axis of the stellar component, $b/a < 0.6$. The two biggest galaxies of the group, ID96526 and
ID96527 (all identification numbers are given at $z=0$), were included among the bar-like galaxies of that study and
had very similar masses and histories. They were assigned to the class A of that sample,
as galaxies that experienced strong interactions with massive objects, which probably led to the formation of a tidally
induced bar-like shape.

It turned out that the two galaxies were accreted quite early by a bigger galaxy, ID96500, which later became the
central galaxy of a cluster (with a total mass of $1.8 \times 10^{14}$ M$_\odot$ at $z=0$). A detailed study of their
evolution revealed that they interacted with each other during the infall and that another nearby galaxy, ID96543, also
interacted with them at this time. This galaxy, although smaller in mass, turned out to have played an
important role in the evolution of the group. The three galaxies, the most massive members of the group, were
accompanied by a number of smaller satellites that did not affect the evolution significantly.

\begin{figure}
\centering
\includegraphics[width=6cm]{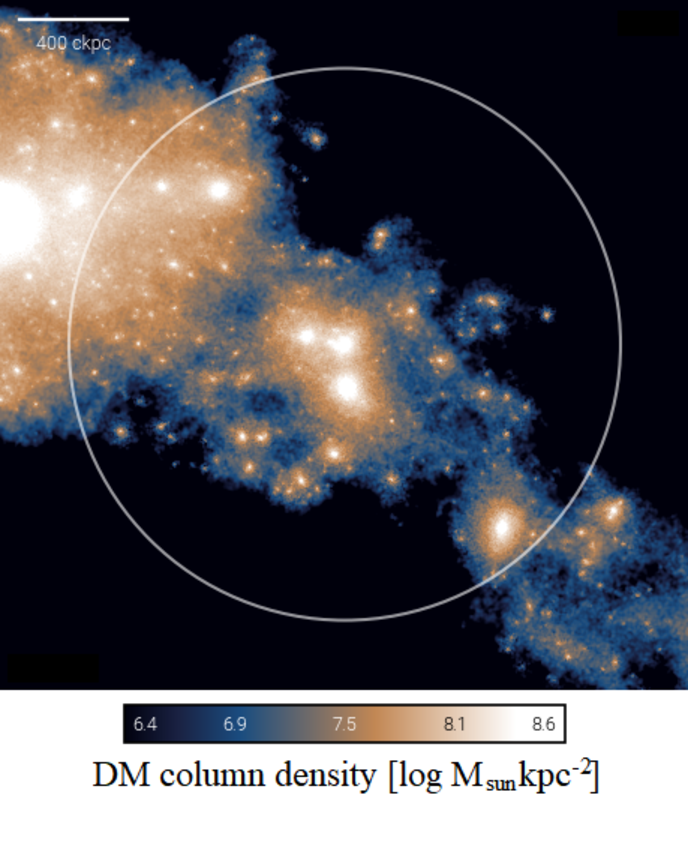}
\includegraphics[width=6cm]{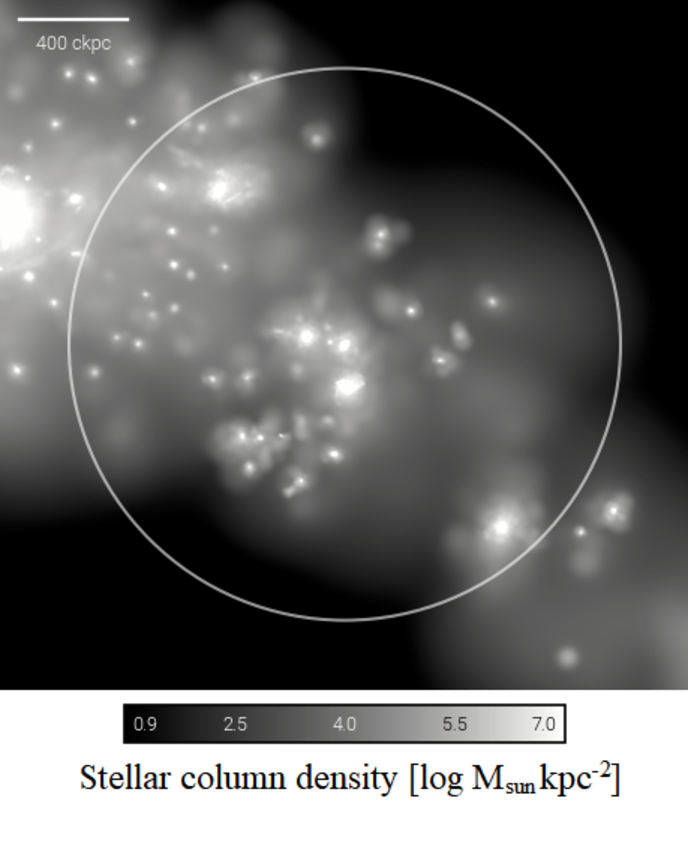}
\includegraphics[width=6cm]{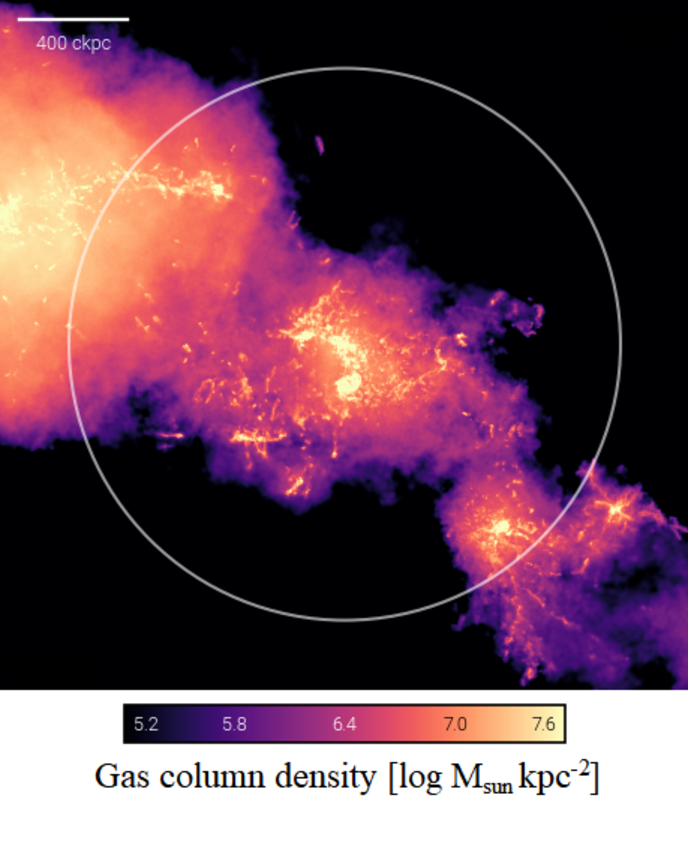}
\caption{Column densities of dark matter, stars, and gas (from top to bottom) in a cutout from the IllustrisTNG100
simulation containing the infalling group of galaxies at $t=4$ Gyr ($z=1.6$). The images are centered on galaxy
ID96543, the galaxy ID96527 is visible to the left of it, and ID96526 is beneath it. The BCG of the forming cluster,
into which the group is infalling, is visible in the upper left part of each panel.}
\label{images}
\end{figure}

Figure~\ref{images} shows the images of the infalling group of galaxies in three components: dark matter, stars, and
gas, in terms of their column densities, at $t=4$ Gyr ($z=1.6$). The three biggest galaxies of the group are well
visible, especially in the dark matter component shown in the upper panel. The brightest cluster galaxy (BCG) of the
forming cluster can be found in the upper left part of each panel.

\begin{figure*}
\centering
\includegraphics[width=18.1cm]{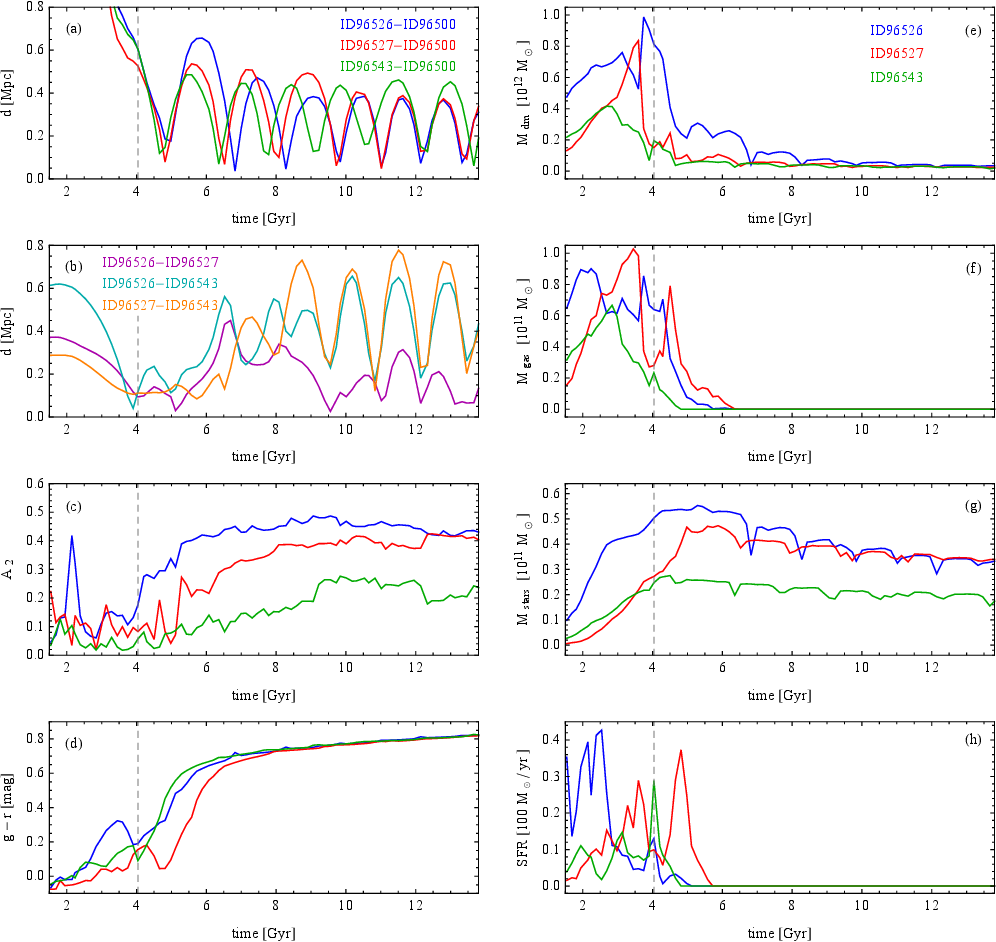}
\caption{Evolution of different properties of the three galaxies over time. Panels (a)-(b) illustrate the orbital
history of the galaxies, with panel (a) showing the distance of each galaxy from the center of the forming cluster and
panel (b) the distance between each pair of galaxies. Panels (e)-(f)-(g) show the evolution of the total mass of each
galaxy in different components, dark matter, gas, and stars, respectively. Panel (c) presents the evolution of the bar
mode $A_2$ in each galaxy, measured within $2 r_{1/2}$. Panels (d) and (h) show the evolution of the color and SFR in the galaxies. The color coding of the lines in panels (c)-(d) and (f)-(h) is the same as in
panel (e). The vertical dashed gray line in all panels indicates the time ($t=4$ Gyr, $z=1.6$) for which the images of
the group are shown in Fig.~\ref{images}.}
\label{evolution}
\end{figure*}

\subsection{Orbital evolution}

The orbital evolution of the interacting galaxies is shown in panels (a) and (b) of Fig.~\ref{evolution}. Panel (a)
plots the distance of each of the three galaxies from the BCG of the cluster. One can see that they all experienced
their first pericenter around $t=4.8$ Gyr and all had seven pericenters around the cluster during their whole lifetime.
In panel (b) I show the evolution of the distance between each pair of the three galaxies. The plot demonstrates that
for about 2 Gyr, that is between $t=3.5$ and $t=5.5$ Gyr, the three galaxies remained within 200 kpc of each other.

The galaxies ID96526 and ID96543 (cyan line) interacted particularly strongly around $t=4$ Gyr, galaxies ID96526 and
ID96527 (magenta line) around $t=5$ Gyr, while ID96527 and ID96543 (orange line) remained within $\sim 100$ kpc from
each other. The first of these strong interactions took place before the galaxies experienced their first pericenter
passage around the cluster, while the second one took place soon afterwards. After infall into the forming cluster, the
group dissolved in the sense that each of the galaxies followed its own orbit around the cluster, although ID96526 and
ID96527 came very close again at $t=9.5$ Gyr and have remained closer to each other than the other pairs.

\begin{figure}
\centering
\hspace{0.3cm}
\includegraphics[width=3.5cm]{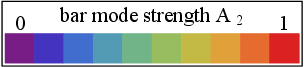}
\includegraphics[width=8.9cm]{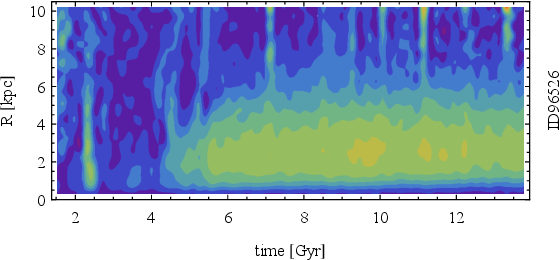} \\
\vspace{0.1cm}
\includegraphics[width=8.9cm]{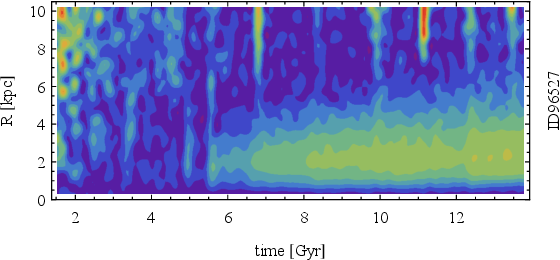} \\
\vspace{0.1cm}
\includegraphics[width=8.9cm]{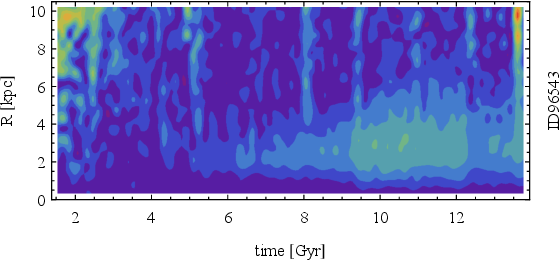}
\caption{Evolution of the profiles of the bar mode, $A_2 (R)$, in the three galaxies over time.}
\label{a2modestime}
\end{figure}

\subsection{Mass evolution}

The evolution of the total mass of the three galaxies in time is shown in panels (e)-(f)-(g) of Fig.~\ref{evolution}
for the dark matter, gas, and stars, respectively. This mass is defined as belonging to the self-bound subhalo and
assigned by the \textsc{subfind} algorithm \citep{Springel2001}. The galaxies ID96526 and ID96527 reached their maximum
dark matter mass on the order of $10^{12}$ M$_\odot$ a little before $t=4$ Gyr, while this mass for ID96543 is about a
factor of two smaller. The maximum gas mass is on the order of $10^{11}$ M$_\odot$ in all three galaxies. The maximum
stellar mass was reached about a gigayear later and is on the order of $5 \times 10^{10}$ M$_\odot$ for ID96526 and
ID96527 and twice as low for ID96543. The maximum total masses of galaxies ID96526,
ID96527, and ID96543 were $(1.12, 0.96, 0.49) \times 10^{12}$ M$_\odot$ at $t = (3.74, 3.59, 2.84)$ Gyr, respectively.
The values correspond to what one would consider to be a normal-sized galaxy at these times.

During the interaction in the group, the galaxies exchanged mass, in particular ID96527 took over some of dark
matter and gas previously bound to ID96543, while ID96526 borrowed some material that previously belonged to ID96527.
After $t=4.5$ Gyr, dark matter and gas were tidally and ram-pressure stripped in all galaxies due to the interaction
with the forming cluster. By the end of the evolution, the galaxies retained only a small fraction of their
original dark matter content and no gas. In particular, the gas was completely lost at the second pericenter around the
cluster for all three galaxies. The stars were less affected, because of their concentration in the galaxy center, but
their mass was also gradually depleted, with stronger dips at pericenters, where the \textsc{subfind} algorithm tends to
underestimate the mass content \citep{Muldrew2011}. The total masses of galaxies ID96526, ID96527, and ID96543 at the
end of the evolution ($z=0$) were $(6.85, 5.88, 3.68) \times 10^{10}$ M$_\odot$, respectively.

\subsection{Evolution of stellar populations}

The interactions of the galaxies in the group, prior to cluster infall, had some bearing on their stellar populations.
The two close encounters between different pairs, at 4 and 5 Gyr, led to the enhancement in their SFR (measured within
two stellar half-mass radii, $2 r_{1/2}$) and the rejuvenation of their stellar populations. As shown in
Fig.~\ref{evolution}(h), galaxies ID96526 and ID96543 interacting at 4 Gyr both display a peak in their SFR at the level of 10
and 30 M$_\odot$ yr$^{-1}$, respectively, around that time, while galaxy ID96527 has an even higher peak at 5 Gyr as a
result of its interaction with ID96526. ID96527 is the only galaxy that still retains a significant amount of
gas around that time. Some of this gas was recently accreted and is rather concentrated, so it is more difficult to strip.

Figure~\ref{evolution}(d) shows the evolution of the $g-r$ color of the galaxies calculated from all stars. Its
behavior reflects what could be inferred from the evolution of the SFR: all galaxies show a drop in color at the time
of their strongest interaction in the group, that is, their stellar population becomes temporarily bluer. However, as
the gas is stripped soon after the cluster infall, the star formation stops and the color increases toward the redder
values. All three galaxies become red with $g-r > 0.6$ mag around $t=6$ Gyr and reach $g-r > 0.8$
mag at the end of the evolution.

\subsection{Morphological evolution}

The most interesting aspect of the interactions of the galaxies in the group and in the cluster is related to their
morphological evolution. It turns out that all the major galaxies of the group formed bars at some stage of their
history. I measured the strength of the bar in terms of the $m=2$ mode of the Fourier decomposition of the surface
density distribution of stellar particles projected along the short axis. It is given by $A_2 (R) = | \Sigma_j m_j
\exp(2 i \theta_j) |/\Sigma_j m_j$, where $\theta_j$ is the azimuthal angle of the $j$th star, $m_j$ is its mass, and
the sum goes up to the number of particles in a given radial bin. Figure~\ref{evolution}(c) shows the evolution of
$A_2$ measured in a single radial bin of radius $R < 2 r_{1/2}$. Assuming the threshold of $A_2 > 0.2$ as
corresponding to the bar formation, it turns out that ID96526 and ID96527 form their bars at 4 and 5 Gyr, respectively.
These times coincide with their strong interactions with other galaxies in the group, which suggests that the formation
of the bars was tidally induced.

At 4 Gyr, ID96526 interacted strongly with ID96543, which induced the bar in ID96526, but left ID96543 rather
unaffected. At 5 Gyr, ID96526 interacted strongly with ID96527, which induced the bar in ID96527, but also enhanced the
bar in ID96526. It is worth noting that ID96527 was already temporarily distorted during its first pericenter
passage around the cluster at $t=4.8$ Gyr, but this interaction did not result in the formation of a stable bar. The
bars then grew steadily, aided probably by the loss of gas due to ram-pressure stripping in the cluster environment. On
the other hand, the bar in ID96543, although seeded by the interactions with the biggest galaxy of the group (ID96526)
and the cluster, grew slowly until 9 Gyr when it crossed $A_2 > 0.2$ probably as a result of the tidal interaction with
the cluster BCG at the fourth pericenter. However, the bar in ID96543 remained much weaker than in the two other
galaxies until the end of the evolution.

A similar enhancement in the bar strength is seen for ID96527 at $t=12$ Gyr, while the bar
in ID96543 was weakened around this time. Both events took place at their respective pericenters on orbits around the
cluster. These are examples of the changes of the bar strength resulting from different orientations of the bar with
respect to the tidal force at the pericenter, an effect described by \citet{Lokas2014}. As demonstrated by their fig.
11, when the tidal torque is oriented so as to speed up the bar, the bar is weakened, while if the torque slows down
the bar, it becomes stronger.

The behavior of the bars in the three galaxies is illustrated in a more complete way in Fig.~\ref{a2modestime}, which
shows the evolution of the full bar mode profile, $A_2 (R, t)$, thus preserving the dependence on the radius. In addition
to the regular enhancement of $A_2$ related to the bar ($R < 6$ kpc), there are a number of other features visible in
these plots. In particular, the short-term enhancement of $A_2$ in the first panel (ID96526) at $t=2.4$ Gyr, also
visible as a strong peak of $A_2$ in Fig.~\ref{evolution}(c) around this time (blue line), is due to the temporary
distortion of this galaxy resulting from a merger. Later on, all galaxies show similar features at larger radii ($R >
6$ kpc), resulting from temporary distortions by tidal forces from another galaxy or the cluster. All of these features
occur at times corresponding to the pericenter passages shown in Fig.~\ref{evolution}(a)-(b).

\section{Discussion}

I presented an analysis of the evolution of three galaxies accreted by a cluster in a small group selected from the
IllustrisTNG100 simulation. Prior to cluster infall, the galaxies interacted strongly, which led to their
morphological transformation. In the two bigger galaxies, the interactions led to the fast formation of strong, tidally
induced bars, while the third, smaller galaxy formed its bar more slowly and the bar was significantly weaker. The bars
grew after the cluster infall of the group, probably aided by the stripping of the gas by the ram pressure in the
intracluster medium. Contrary to what is observed in nearby clusters, the group accretion took place very early in the
evolution of the cluster, with the first pericenter passage around $z=1.3$ ($t=4.8$ Gyr). Interestingly,
the imprint of the interactions in the group, in terms of the tidally induced bars, was preserved until the end of the
simulation corresponding to the present time, in spite of the strong evolution in the cluster, including seven
pericenter passages.

The process of the creation of tidally induced bars during the flybys in the group and their subsequent evolution in
the cluster can be viewed as another scenario for their formation, combining idealized two-body flybys
\citep{Lokas2018} and interactions of a single galaxy with a cluster \citep{Lokas2016}. Otherwise, it closely resembles
such events described in previous studies performed using controlled or cosmological simulations \citep{Lokas2014,
Lokas2020b, Lokas2021}. Although cosmological simulations provide a more realistic environment to study the evolution,
their disadvantage is that one cannot claim unequivocally that the bars did indeed result from tidal interactions,
because of the lack of a control experiment where the same galaxy would evolve without taking part in the encounter.

Another source of uncertainty is related to the possible effects of other processes occurring simultaneously with the
bar formation. For example, in the case of galaxy ID96526 studied here, the formation of the bar coincides with an event
of strong kinetic feedback from its AGN. A significant energy released at this time led to the heating of the inner gas
and its removal from the central few kiloparsecs of the galaxy. The resulting lower gas content may have facilitated the
formation of the bar, as described by \citet{Athanassoula2013} and \citet{Lokas2020a, Lokas2022}.

The presented scenario of the evolution of galaxies in a small group quite early in the cosmological timescale offers
an interesting possibility located in between the secular evolution in isolation and the strong evolution in clusters.
The specific environment of the group is where the galaxies come together close enough to
interact but at the same time their relative velocities are not yet as high as in clusters, making the interactions
very effective in terms of the strength of morphological transformation. The time of the interaction is also quite
special, because it happened when the galaxies were already fully formed, having reached the dark and stellar masses
typical of normal-sized present-day galaxies. Additional analyses of cosmological simulations and observations
are needed to determine if this scenario is an exception or a common occurrence at intermediate redshifts.

\begin{acknowledgements}
I am grateful to the IllustrisTNG team for making their simulations publicly available and to the anonymous
referee for useful comments.
Computations for this article have been performed in part using the computer cluster at
CAMK PAN.
\end{acknowledgements}

\end{document}